\documentstyle[aps,pre,twocolumn,epsf]{revtex}

\newcommand{\equ}[1]{(\protect\ref{#1})}

\begin{document} 

\draft

\wideabs{

\title{Fractal properties of clusters of colloidal magnetic particles}

\author{R. Pastor-Satorras}

\address{Department of Earth, Atmospheric, and Planetary Sciences\\
Massachusetts Institute of Technology, Cambridge, Massachusetts
02139, USA}
\author{J. M. Rub\'{\i}}
\address{Departament de F\'{\i}sica Fonamental, Facultat de
F\'{\i}sica\\ 
Universitat de Barcelona, Diagonal 647, 08028 Barcelona, Spain}

\maketitle

\pacs{{\em Keywords:} Magnetic particles --- ferrofluids ---
cluster aggregation --- dipolar interactions}

\begin{abstract}

We have studied the properties of clusters of colloidal magnetic
particles generated from a 2D 
aggregation model with dipolar interparticle 
interactions. Particles diffuse off-lattice, experiencing dipolar
interations with the already attached particles until either they
stick to the cluster or wander far away and are removed.
Our results are interpreted in terms of a 
fractal dimension that is a 
monotonically decreasing function of the temperature, varying between
a definite 
value close to $1$ at $T=0$, and the limit $T\to\infty$,
corresponding to free  diffusion-limited aggregation.
By analyzing orientational correlation functions, an ordered state is
found at low temperatures; this state is destroyed by the fractal
disorder generated at high $T$.
Our study 
could be relevant in understanding aggregation of dipolar colloids and
phase transitions in Langmuir monolayers.

\end{abstract}

}

\section{Introduction}

Fractal growth phenomena \cite{vicsek92} have been a particularly active
field of physics in the last decade, due to the potential applications
to many disciplines, in particular to the physics of colloids. In the
context of computer models, the most noteworthy are the
diffusion-limited 
aggregation (DLA) model  \cite{witten81}
and the cluster-cluster aggregation model \cite{meakin83}, which
indeed 
can describe the fractal structure of colloidal aggregates
\cite{weitz84}. 

Most often the above mentioned models assume  very short range
interparticle interactions; usually a hard-core potential plus
an infinite well on the surface of the particles. They are therefore
appropriate to describe aggregation  with
interactions strongly decaying with distance. However they fail to
represent aggregation in the presence of long-range  forces.
Some examples of such kind of processes are the aggregation of
particles subject to {\em dipolar forces}. Actual experiments have been
conducted in ferrofluids and the so-called magnetic holes
\cite{skjeltorp83}; part of their interest lies in the extreme
simplicity of the experimental setup, which can be easily performed
with magnetized microspheres \cite{helgesen91}. On the other hand,
there is a considerable interest in the study of phase transitions in
Langmuir monolayers \cite{mohwald93}. Far from equilibrium, the
condensed phase grows at the expense of the liquid phase, forming
clusters of different shapes. The phospholipids constituting the
Langmuir monolayer experience repulsive dipolar interactions, which
must play a major role in determining the morphology of the
condensed aggregates.

Some authors have extended the classic  models in order to
take interparticle interactions into account. The modifications
proposed so far are either 
deterministic \cite{block91,eriksson89} or random
\cite{ansell85,meakin89,pastor95,indiveri96}, using Monte Carlo or
Langevin dynamics methods. However, because the implementation of long
range interactions  is extremely time-consuming,
only modest-sized aggregates can be grown (up to 128 particles in
\cite{eriksson89,mors87,helgesen88}). Therefore the results do not
allow categorical conclusions to be made about the fractal properties
of the dipolar clusters. 

Our purpose in this paper is to extend the particle-cluster
aggregation model in order to 
include fully anisotropic attractive dipolar interparticle
interactions.  
Our algorithm is considerably
fast, allowing us to generate clusters up to $10000$ particles at zero
temperature, in a reasonably short time.
In Section~II we describe the technical details of the algorithm.
Section~III discusses the evolution as a function of temperature of
the fractal dimension and the order induced on the orientation of the
dipoles by their reciprocal interactions.
Our conclusions are discussed finally in Section~IV.

\section{Cluster formation algorithm}

We consider the two-dimensional aggregation
process of magnetic particles of diameter $d$ and magnetic moment
$\vec{\mu} = \mu \vec{u}$, with $\mu$ being
the magnetic moment strength and $\vec{u}$ a unit vector
oriented along its direction. The dipolar energy between two particles
$i$ and $j$, located at the positions $\vec{r}_i$ and $\vec{r}_j$,
respectively, is ${\cal E}_{ij}= \mu^2 E_{ij}$, $E_{ij}$ being the
dimensionless dipolar energy 
\begin{equation}
E_{ij}= \left\{ \vec{u}_i \cdot \vec{u}_j - 3(\vec{u}_i \cdot
\vec{r}_{ij}) (\vec{u}_j \cdot \vec{r}_{ij}) / r_{ij}^2
\right\} / r_{ij}^3,
\label{energy}
\end{equation}
and $\vec{r}_{ij} = \vec{r}_i - \vec{r}_j$.

The simulation starts with a seed particle located 
at the origin of coordinates, bearing a randomly oriented tridimensional
vector $\vec{u}_1$, parallel to the plane of growth. The
following particles are released from a random  position on a
circle of radius $R_{\mbox{\rm\scriptsize in}}$ centered on the seed. The
particles have assigned a vector $\vec{u}_i$
oriented at random. Each particle undergoes a random walk until it either
contacts the cluster or moves away from the origin a distance greater than
$R_{\mbox{\rm\scriptsize out}}$. In this case, the particle is removed and a
new one is released from the circle surrounding the seed. 
We have used the values $R_{\mbox{\rm\scriptsize
in}} = 2 R_{\mbox{\rm\scriptsize max}} - 5$ and
$R_{\mbox{\rm\scriptsize out}} = 2 R_{\mbox{\rm\scriptsize max}}$, where
$R_{\mbox{\rm\scriptsize max}}$ is the maximum radius of the cluster. 
The random walk experienced by the incoming particles is affected by the
interactions exerted by the particles already attached to the cluster. We
have taken this fact into account by using a Metropolis algorithm inspired by
Refs.~\cite{mors87,helgesen88}. 
Suppose that the cluster is composed by $n-1$ particles, placed at the
points $\vec{r}_i$, $i=1,\ldots,n-1$. At some time $t$
the incoming particle  occupies the position
$\vec{r}_n$ and has an interacting energy $E=\sum_{i=1}^{n-1} E_{ni}$.
At time $t+1$  we compute a new position
${\vec{r}_n}^*$; the particle arrives there
by means of jump of length $d$ in a direction
chosen at random. The movement to 
${\vec{r}_n}^*$ is performed rigidly, without changing
$\vec{u}_n$. The energy experienced in the new position
is $E^*$, and the total change in the energy
due to the movement is $\Delta E = E^* - E$. If $\Delta E < 0$, then
the movement is accepted; if $\Delta E > 0$, we compute the
quantity $p=\exp(-\Delta E / T_r)$, 
where
\begin{equation}
T_r=\frac{d^3 k_{\mbox{\rm\scriptsize B}} T}{\mu^2}.
\label{parametro_k}
\end{equation}
($T_r$ is a {\em reduced temperature}, related to the intensity of the
interaction and the actual temperature $T$.) In this latter 
case, the movement is accepted with probability $p$. 
After every
accepted movement, the moment of the random walker is oriented along
the  direction of the total field on its position.
This fact indeed assumes that the relaxation time for the orientation of
particles is very short in comparison with the movement of the center of
mass.
The particle sticks to the cluster when it overlaps one or more particles
already incorporated. After sticking, the newly attached particle
undergoes one last relaxation.

\section{Fractal properties of the clusters}

\begin{figure}
\epsfxsize=8truecm    \centerline{\epsfbox{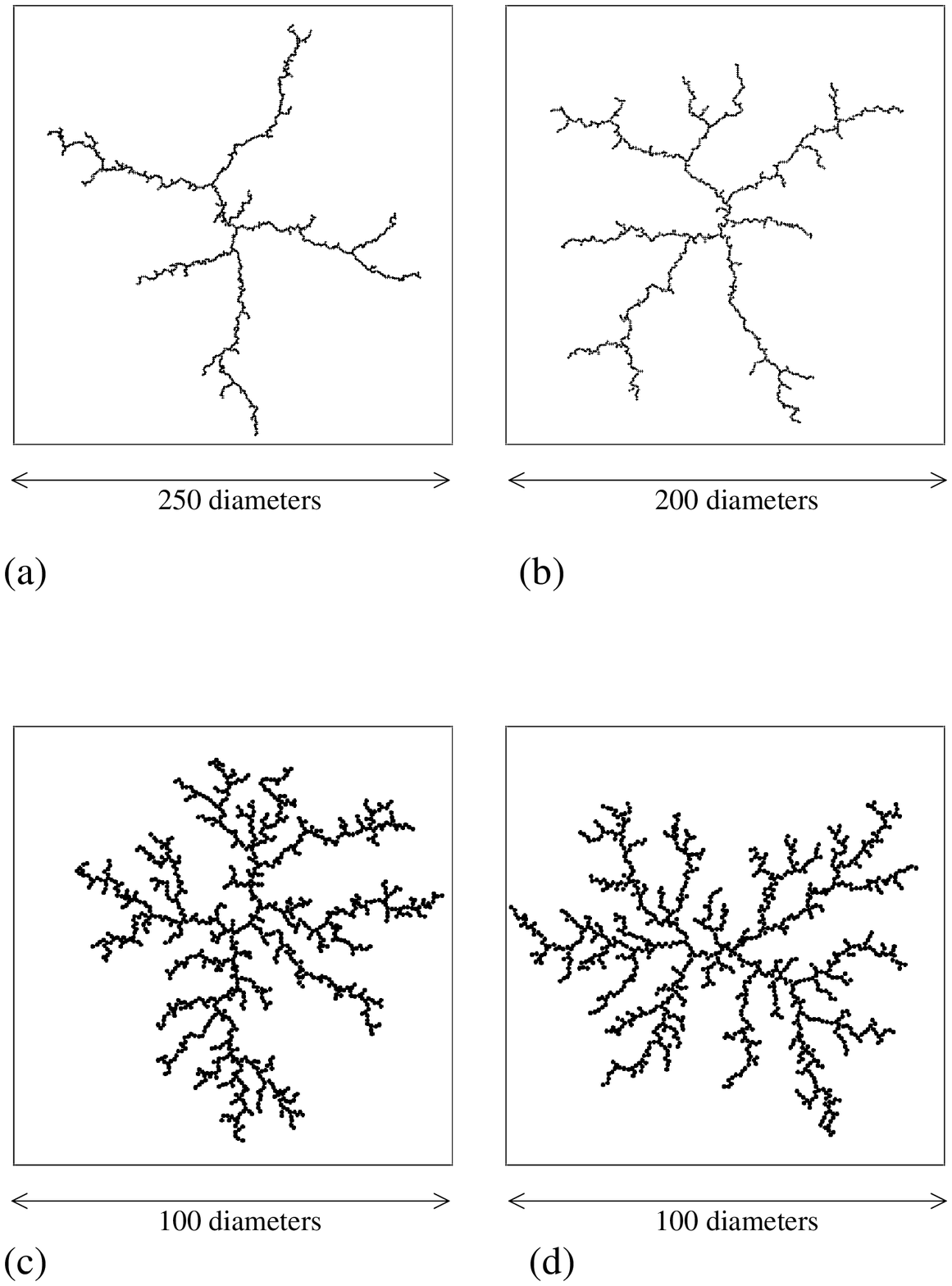}}
\caption{Typical 2D dipolar clusters of
$1000$ particles, generated for several values of $T_r$. (a) $T_r=0$,
$D=1.20\pm0.02$. (b) $T_r=10^{-3}$, $D=1.35\pm0.04$. (c) $T_r=10$,
$D=1.74\pm0.02$. (c) Pure DLA cluster, grown in the limit $T_r \to
\infty$, $D=1.71\pm0.01$.}
\label{big1}
\end{figure}

The purpose of this Section is to analyze the fractal properties of
the clusters generated using the prescription outlined above. In
Fig.\ref{big1} we have represented typical clusters of $1000$
particles, grown at four  
different values of $T_r$. The effects of temperature can be seen by
comparing with a pure non-interacting DLA cluster, as shown in
Fig.\ref{big1}(d). At low $T_r$, Fig.\ref{big1}(a), the dipolar
clusters have a lesser 
branched and more open structure, that is to say, they have a smaller
ratio of bifurcation. Even though the clusters may
seem to be anisotropic, they still posses spherical symmetry: When
collapsing an ensemble of clusters at the same $T_r$, we recover a
perfectly symmetric structure. When increasing the temperature, the
branching of the clusters increases correspondingly; for $T_r=10$ (the
largest value simulated),
Fig.\ref{big1}(c), the 
clusters are completely indistinguishable from true DLA.

%FIGURE 1

In order to quantify this temperature dependence, we have computed the
fractal dimension $D$ of the clusters, determined from a log-log plot
of the radius of gyration as a function of the number
of particles $N$ \cite{vicsek92},
\begin{equation}
R_g(N)=\left( \frac{1}{N} \sum_{i=1}^{N} (\vec{r}_i-
\vec{r}_{\mbox{\rm\scriptsize c.m.}})^2 \right) ^{1/2} \sim N^{1/D}.
\end{equation}
The algorithm was checked by computing the dimension of an ensemble of
$100$ clusters of $1000$ particles, generated in the  limit
$T_r\to\infty$ 
(pure DLA). The value computed was $D=1.71\pm0.01$; we 
recover, within the error bars, the well-known result $D=1.715\pm0.004$
\cite{tolman89}.

\begin{figure}
\epsfxsize=8truecm    \centerline{\epsfbox{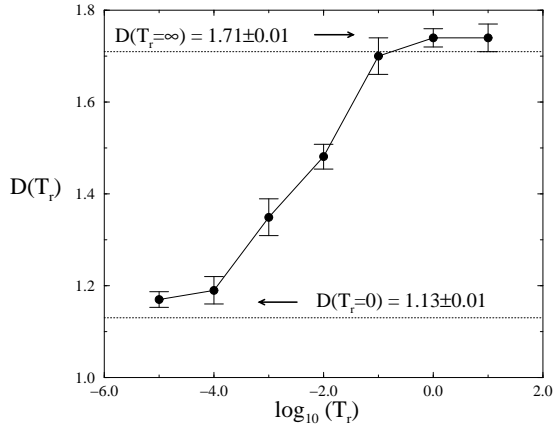}}
\caption{Fractal dimension $D$ as a function of $T_r$ for dipolar
clusters.}
\label{dimension}
\end{figure}

Fig.\ref{dimension} shows a plot of $D$ as a function of $T_r$ for
the whole range 
of values analyzed. It seems to exhibit a {\em smooth} increasing behavior
when increasing $T_r$, the values of $D$ ranging between the limits
corresponding to $T_r=0$ (aggregation with dipolar forces of infinite
strength or zero temperature) with $D=1.13\pm0.01$ and
$T_r=\infty$ (aggregation with no interactions or infinite
temperature), with $D=1.71\pm0.01$. We can compare our results with the
colloidal aggregation 
experiment described in \cite{helgesen88}, where magnetic particles
were employed for which $T_r^{-1} \simeq 1360$ at room temperature. In our
simulations, the 
fractal dimension obtained for clusters grown at $T_r=10^{-3}$ is
$D=1.35\pm0.04$, a value clearly different from the one computed at
$T_r=0$.

%FIGURE 2

The shape of Fig.\ref{dimension} can be explained as follows. Given
the 
expression~\equ{energy} for the dipolar energy, it is energetically
favorable for an incoming particle to stick at the tip of a branch,
contributing thus to its growth, rather than sticking on any side and 
splitting it. At low temperatures, therefore, the most probable
scenario is that of a cluster with a very small branching ratio and a
low fractal dimension. This in consistent with the fact that, at
$T_r=0$, the fractal dimension seems to decrease when increasing $N$;
we cannot even reject the possibility of a dimension equal to $1$ in
the limit $N\to\infty$. On the other hand, at large temperature
the growth of a given branch and its split are equally likely 
events. The effect of dipolar interactions is overcome by thermal 
disorder and we  recover the original DLA model with no interactions. At 
intermediate values of $T_r$, there would be a competition between
growth and splitting, ruled by the thermal disorder as well as the
dipolar interactions. We should then expect a continuous change
in the  geometry of the cluster and, therefore, a smooth dependence of
$D$ on the  temperature. 

The fractal dimension we have considered so far is a {\em macroscopic}
property of dipolar aggregates. 
However, their composing particles also  bear a rigid magnetic dipole,
which confers a {\em microscopic structure} to the clusters. In order
to obtain information about this structure, we have analyzed the
{\em relative orientation} between pairs of dipoles. To this end, we
define the function $g_T(\theta)$ as the probability density that the
relative angle formed by the directions of a pair of dipoles randomly
chosen from a cluster at temperature $T_r$ is included in the interval
$\left[\theta,\theta+d\theta\right]$; $\theta$ is defined to be
normalized to the interval $\left[0,\pi\right]$.
In practice, if  $n(T_r,\theta)$ is the number of pairs with a
relative angle between $\theta$ and $\theta+\Delta\theta$, for a fixed
angular increment $\Delta\theta$, then we have
\begin{equation}
g_T(\theta)=\frac{2}{N(N-1)} \frac{1}{\Delta\theta} n(T_r,\theta),
\label{o_p_d}
\end{equation}
$N$ being the number of particles in the cluster.
$g_T$ is a measure of the order of the dipoles
on the cluster. In a completely disordered distribution, all relative
orientations are equally probable and, therefore, we have
$g(\theta)=1/\pi$. On the other hand, in a distribution in
which all the dipoles point in the same direction (for  example, in
the presence of a strong magnetic field) every pair forms a relative
angle of zero radians, and thus $g(\theta)=\delta(\theta)$.

\begin{figure}
\epsfxsize=8truecm    \centerline{\epsfbox{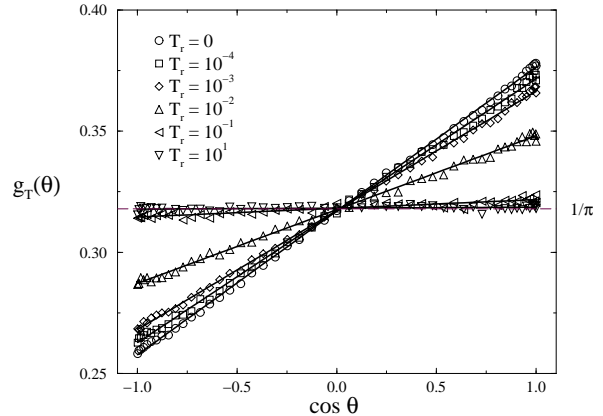}}
\caption{Orientational correlation $g_T(\theta)$ as a
function of 
$\cos \theta$, computed for dipolar
clusters at different 
values of $T_r$. At high temperatures the slope is almost zero.  Solid
lines are least-squares fittings.}
\label{correlations}
\end{figure}

Fig.\ref{correlations} depicts $g_T(\theta)$ computed from several
ensembles of 
clusters  grown at different values of $T_r$,
between $0$ and $10$. 

%FIGURE 3

Numerically we observe that 
$g_T(\theta)$ fits extremely well to the function
\begin{equation}
g_T(\theta)=a+b(T_r)\cos\theta.
\label{coseno}
\end{equation}
The normalization condition of $g_T$ implies obviously $a=1/\pi$, as
observed. On the other hand,  $b(T_r)$ seems to be a decreasing
function of temperature, between the limits $b(0) = 0.0594\pm0.0005$
and $b(\infty) \sim 0$.
The 
slope $b(T_r)$ can be seen as a measure of the degree of internal
order in the orientation of the dipoles in the cluster. In the low
temperature limit (in the absence of any thermal disorder) the
orientation of the dipoles is strongly correlated, and this fact is
reflected in a nonzero slope $b$. As temperature raises, $g_T(\theta)$
decreases continuously. In the high temperature limit dipoles show a
large disorder, imposed by the intrinsic fractal geometry, and the
function  $g_T$ is almost flat (all relative orientations are equally
probable).   Once again, the apparently continuous
variation of  $g_T$  hints towards a smooth dependence of the geometry
of the clusters on the temperature.

\section{Conclusions}

We have investigated the effects of dipolar 
interactions in particle-cluster aggregation in two dimensions.
The relevant parameter in the model is the dimensionless temperature
$T_r$ defined in~\equ{parametro_k}, which relates the real temperature
$T$ and the strength of the magnetic interactions $\mu$.
At low temperatures we observe clusters with a less branched and more open
structure than free DLA, and correspondingly, a fractal dimension
close to $1$. This  is an effect of the dipolar interaction
on the local growth-site probability distribution
\cite{vicsek92}. The growth of a given branch is a more likely event than
its split, resulting in an enhancement of the screening of the
inner regions.
At high temperatures  the fractal dimension
raises its value until it reaches the limit of free DLA.  The plot of
$D$ as a function of $T_r$, Fig.\ref{dimension}, seems to 
show a smooth behaviour, resulting from the competition
between dipolar attractive forces and thermal disorder.
The internal structure of the clusters, probed through the function
$g_T$, seems also to show a smooth transition between an ordered state
at low temperature, with long range correlations between dipoles, and
a disordered state at high temperature, in 
which all relative orientations are equally
probable. This loss of order is explained as the effect  of the
geometrical disorder induced 
by the fractal character of the clusters.
Our results could be relevant to understand the processes of cluster
aggregation in dipolar colloids. Even though our result are relative to
attractive dipolar interactions, they could also be applied to study
cluster growth in Langmuir monolayers; the similarity with
experimental clusters has already been pointed out in
Ref.\cite{indiveri96}, in the particular case of {\em isotropic}
dipolar interactions (potential decaying with distance as $r^{-3}$).

As a final remark, we would point out that, 
even though all our results are consistent with a geometry continuously
varying   with the temperature, they are also compatible with a
different scenario, in which  
there is sharp crossover between the low temperature dimension $D_0 
\simeq 1.13$ and the high temperature dimension $D_\infty \simeq 1.71$. 
From Fig.\ref{dimension}, the crossover temperature could be
estimated to be $T_c \simeq 10^{-3.5} \sim 3 \times 10^{-4}$. The
presence of the sharp fall 
in $D$ would then be smoothed out by finite-size effects, unavoidable for
the cluster sizes we are considering. Further work should be done in
order to elucidate this possibility, especially by simulating
aggregates larger than those actually available with our present
computer resources.

\acknowledgments

RPS benefited from a scholarship  grant from the
Ministerio de Educaci\'on y Cultura (Spain). JMR acknowledges
financial support by CICyT (Spain), Grant no. PB92-0895.

%
%       References
%

\end{document}